\documentclass[a4paper]{article}

\usepackage{icrc2013}
 \usepackage[english]{babel}

\title{The MAGIC telescopes DAQ software and the on-the-fly online analysis client}

\shorttitle{MAGIC telescopes DAQ and online analysis}

\authors{
Diego Tescaro$^{1,2}$,
Alicia L\'opez-Oramas$^{3}$,
Abelardo Moralejo$^{3}$,
Daniel Mazin$^{4}$,
Daniela Hadasch$^{5}$
for the MAGIC Collaboration.
}

\afiliations{
$^1$  Inst. de Astrof«õsica de Canarias, E-38200 La Laguna, Tenerife, Spain \\
$^2$  Universidad de La Laguna (ULL), Dept. Astrof«õsica, E-38206 La Laguna, Tenerife, Spain \\
$^3$ IFAE, Edifici Cn., Campus UAB, E-08193 Bellaterra, Spain \\
$^4$ Max-Planck-Institut fur Physik, D-80805 Munchen, Germany\\
$^5$ Institut de Ci\`encies de l'Espai (IEEC-CSIC), E-08193 Bellaterra, Spain
}

\email{dtescaro@iac.es}

\abstract{
In this contribution we describe the design of the Data AcQuisition (DAQ) and online analysis software of the MAGIC telescopes after the 2012 upgrade. 
Although the final stereo trigger requires coincidence between the two telescopes, the actual data acquisition is performed independently, producing two separate data streams. 
Events are first readout and built from the front-end electronics and then stored in the DAQs' internal ring buffer for further processing: pre-calibration and signal extraction. 
The pixel signals, previously used only for data quality monitoring, are now also sent ``on-the-fly'' to the centralized online analysis program MOLA, which 
acts as a single client for the two DAQ data streams, and uses this information to provide preliminary high level analysis results. 
The integrated DAQ and online analysis programs allows an immediate feedback in case of a rapid $\gamma$-ray flare of the pointed astrophysical source.}


\begin{document}
\maketitle

\section{Introduction} 

MAGIC is a stereoscopic system of two Cherenkov telescopes for the detection of gamma rays of energy above ~50 GeV up to few TeV. 
In 2011 and 2012 the MAGIC telescopes underwent a major upgrade \cite{bib:upgrICRC}. 
The main changes affected the camera detector of the MAGIC-I telescope, which was replaced, and the readout system of both MAGIC telescopes, which were upgraded in their core part: the digitization electronics \cite{bib:drs4IEEE}.

Cherenkov telescopes require very fast readout acquisition systems \cite{bib:icrc2009readout}. 
The light flashes (Cherenkov photons) produced by the energetic cosmic-ray showers in the atmosphere are very short in time, of the order of few ns only. 
When an event triggers the MAGIC telescopes a 30~ns long snapshot of the light arriving to the camera detector of the telescopes is taken and stored as raw data event.
To do so, the phototube pixels of the camera detector catch the Cherenkov photons impinging the photocathode and convert them in photo-electrons (hereafter: phe). 
These analog signals go through the early stages of the electronic chain and  are finally convoyed to the readout system, appointed to perform the pulse signal digitization.
Currently 1039 pixels per telescope camera are sampled at the frequency of 2~GSamples/s.
This allows a good reconstruction of a typical 2-3~ns wide pulse signal.
After sampling, the digital information is hold temporally in the readout FIFO memories, until it is readout from the front-end electronics by the DAQ software (section \ref{daq}).
The acquisition in each telescope is completely independent, except for the stereo trigger signal which is delivered simultaneously to both telescopes.

During data acquisition, the DAQ software extracts from the data also some higher level analysis quantities, like the pulse integral and arrival time of the signal in each pixel of the camera.
This was initially meant only for online data checking, in order to immediately spot eventual problems in the hardware.
No higher level analyses are possible at DAQ level, given the large overhead that it would required in terms of computing power\footnote{To obtain high level analysis results is also beyond the scope of a data acquisition system.}.
Nevertheless, these already available pixel-wise informations (signal intensity and arrival time) can be sent, with limited effort, through the local network to a second program running on an independent computer.
This program (section \ref{mola}) has the specific task of analyze the data up to the highest possible level (physics results).
No extra storage of data is requited, since the events are processed on-the-fly and immediately discarded once the relevant informations are adsorbed. 

\section{The data acquisition software} \label{daq}

\emph{Domino4Readout} is the current MAGIC DAQ program.
It is based on the previous version, \emph{DominoReadout} \cite{bib:dtthesis}, with some major updates and improvements.
It is written in C/C++ language with the support of the \emph{pthread} POSIX library for parallel processing.
Two identical copies (except for the configuration files) of this software run on two dedicated server computers, each one appointed to the data acquisition of one of the MAGIC telescopes.

\begin{figure}
\centering
\includegraphics[width=0.4\textwidth]{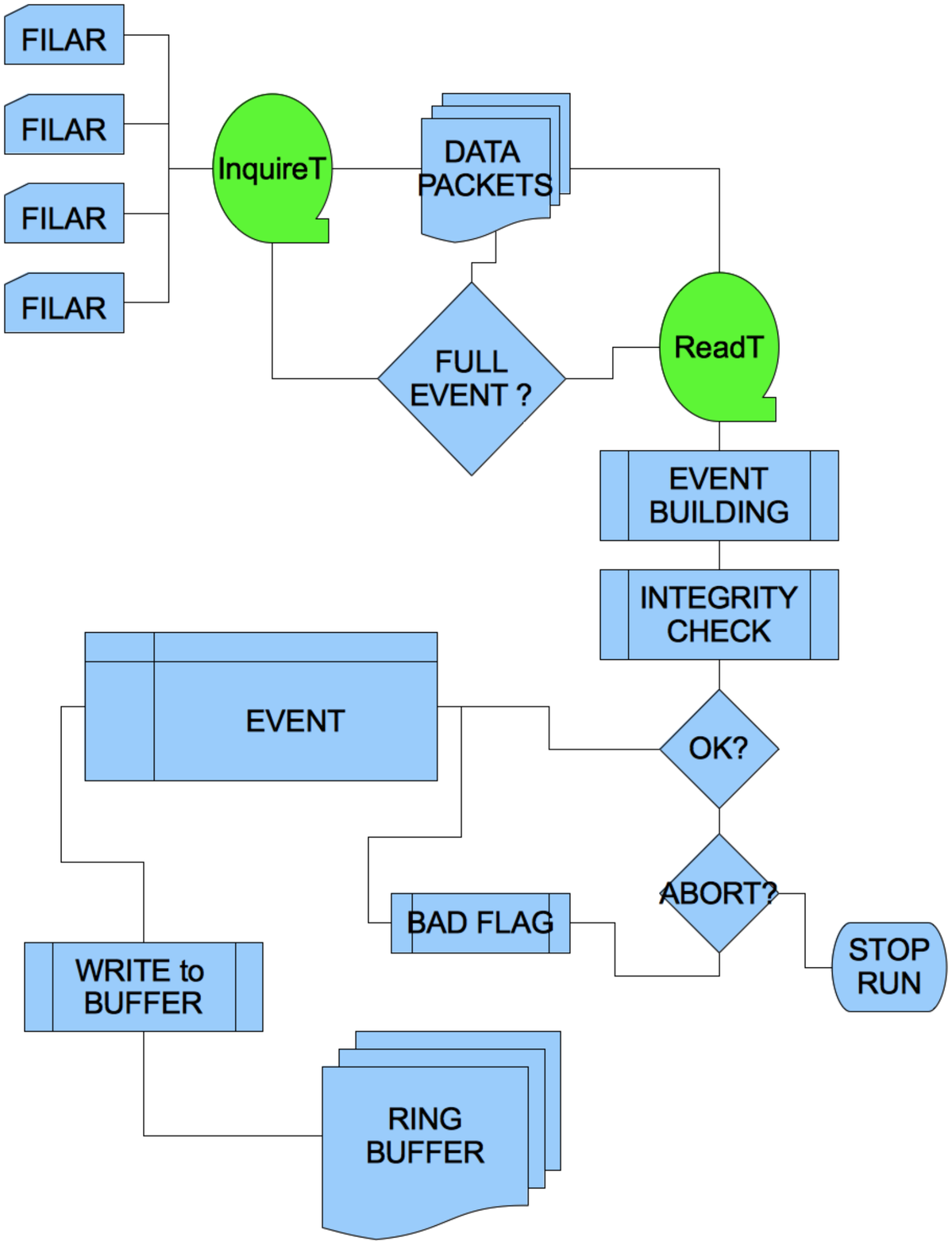}
\caption{Scheme of the DAQ \emph{reading} thread. Note that the event building and the event inquiring are performed in parallel by two independent threads.}
\label{simp_fig1}
\end{figure}

From the DAQ point of view the acquisition data flow starts at the receiver cards installed in the DAQ computer \cite{bib:icrc2009readout}
and finishes at the SATA raid disk system (fiber channel linked), where the raw data are finally stored for further offline processing and archiving (see \cite{bib:MARS}).

The main tasks of the DAQ software can be divided in two categories: \emph{core} and \emph{auxiliary}.
The core tasks are the steps which every event has to undergo, whereas the auxiliary tasks regard mainly the steering of the program and the complementary functions of the DAQ.
The core stages can be summarized as follow, and must be performed in this specific order:
\begin{itemize}
\item Data packets collection from the data link of each readout board (acquisition starts only when at least the information for one complete event is available).
\item Event building, with reordering of the sample information and creation of the event header.
\item First-level integrity checks: consistency of control words, consistency of the local trigger number, etc..
\item Pre-processing of the data samples: cells offset calibration and time lapse corrections (see text).
\item Basic pixel-wise data analysis: pulse intensity and arrival time calculation.
\item Second-level integrity checks: signal check consistency in calibration events.
\item Storage of the event to disk in raw data files.
\end{itemize}

The auxiliary functions can instead be performed with no specific order constraint. 
Please find below a list of the most relevant taks:
\begin{itemize}
\item Console input/output and manual commanding.
\item Remote central control reporting and commanding.
\item DAQ state and operation log.
\item Data-check values saving.
\item Data streaming of the pixel wise information to the online analysis client.
\end{itemize}
\begin{figure}
\centering
\includegraphics[width=0.4\textwidth]{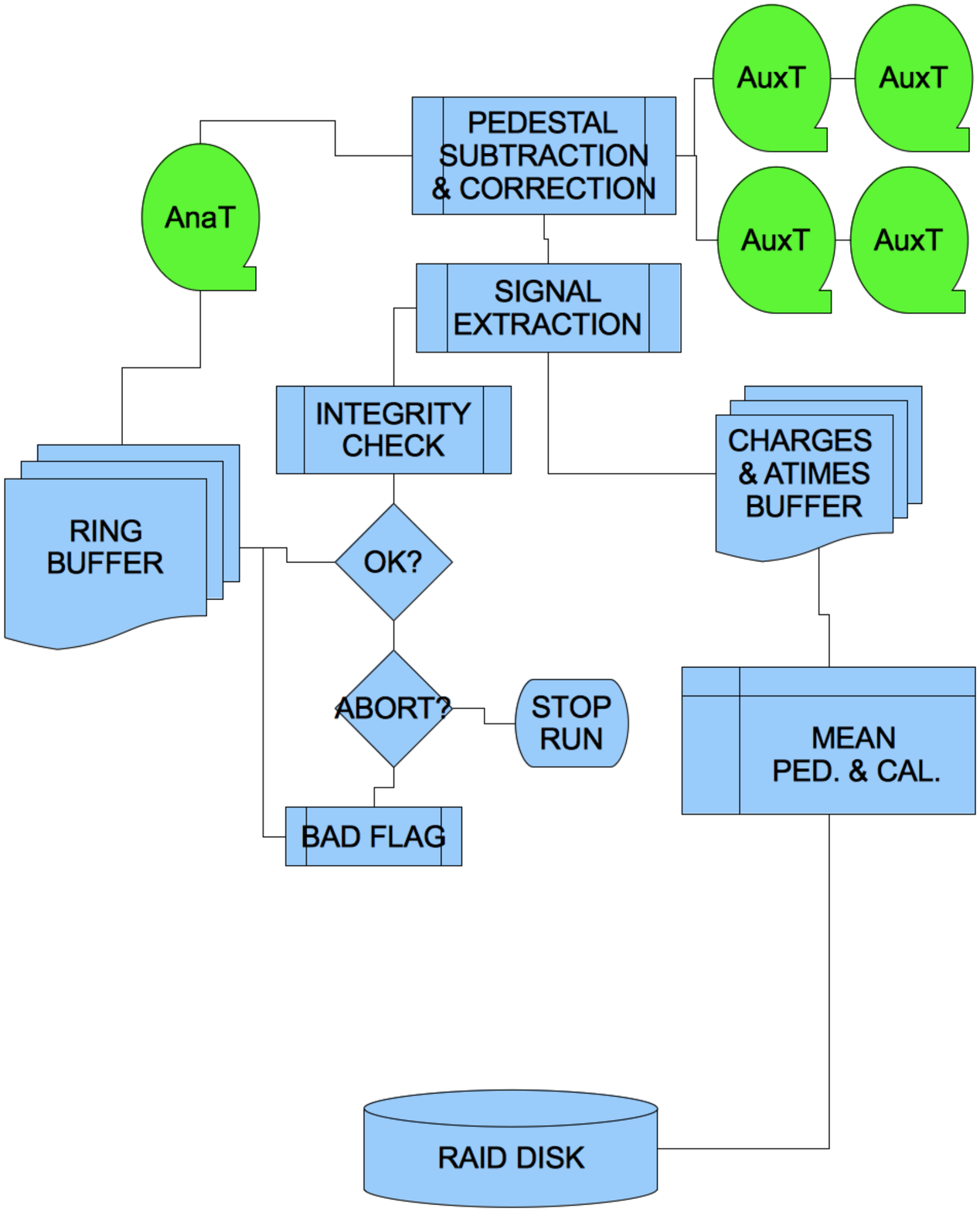}
\caption{Scheme of the DAQ \emph{analyzing} threads. Different chunks of the same events are analyzed in parallel by four auxiliary threads.}
\label{simp_fig2}
\end{figure}

\begin{figure}
\centering
\includegraphics[width=0.4\textwidth]{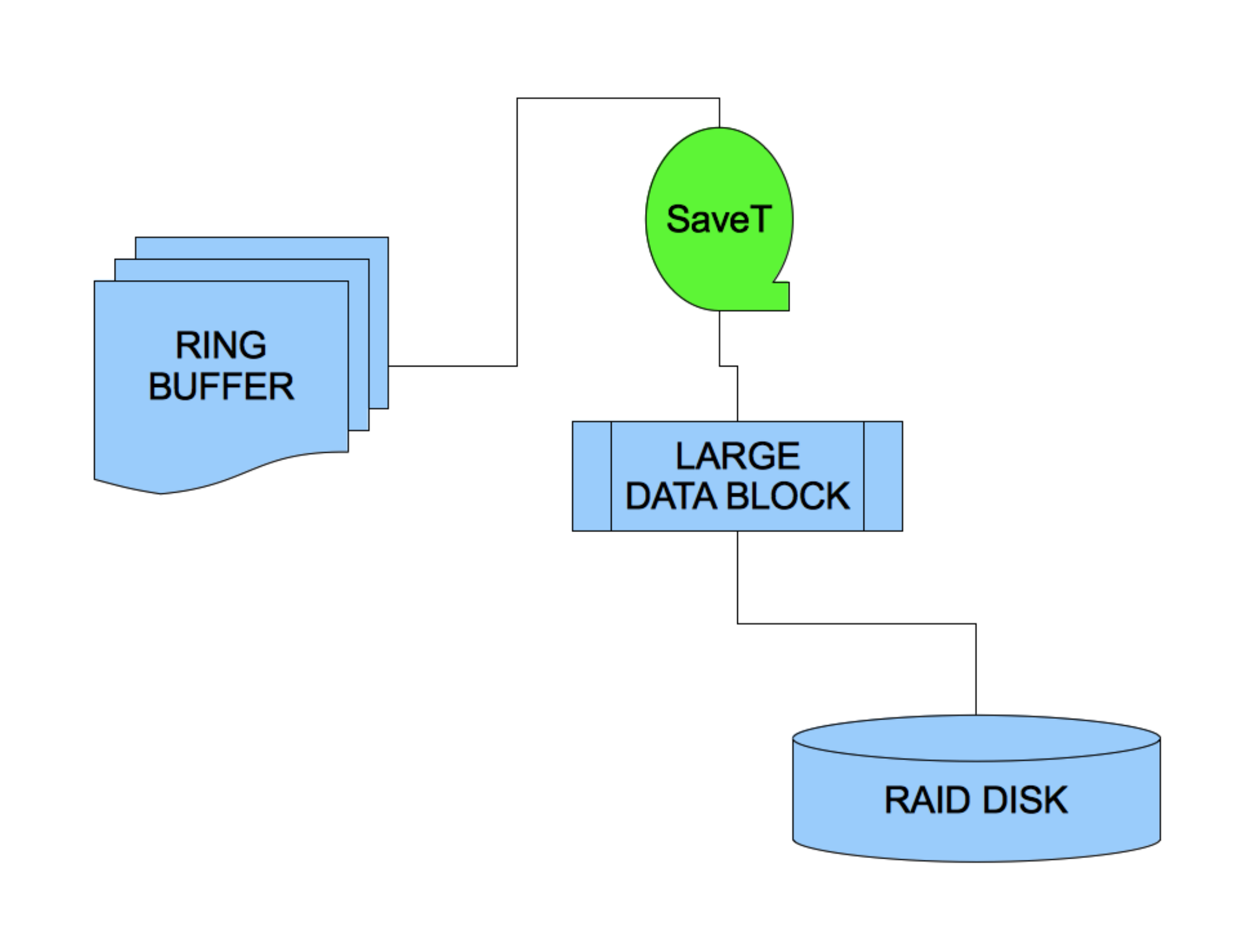}
\caption{Scheme of the DAQ \emph{saving} thread. Data are grouped in larger data blocks of several events in order to optimize the disk data~I/0 performance.}
\label{simp_fig3}
\end{figure}

The major upgrades of data acquisition software with respect to the previous version regard the data packet interpretation, being the format changed with the use of the new DRS4 sampler, and the pre-processing of the data samples values, which is now different with respect to treatment reserved to the previous DRS2 chip.
The pre-processing \cite{bib:perfNIMA} of the samples consists now of two steps: the \emph{mean cell offset} regularization (a sort of internal calibration of the pedestal of the single sampling unit) and the \emph{time lapse correction} (being the offset level dependent on the time passed from the previous readout).
The former is a major correction, absolutely required to lower the pedestal noise by an order of magnitude whereas the latter is a second order correction but still required in order not to loose sensitivity for low signals.
A third correction step is also required, the \emph{arrival time calibration}, but it is not performed online since not as crucial as the other two.
A scheme of the key threads of the DAQ are shown in figures \ref{simp_fig1}, \ref{simp_fig2} and \ref{simp_fig3}.

For the DAQ programming, the most challenging point was the application of the online time lapse baseline correction.
This data correction requires the knowledge of the time passed between two subsequent hardware readout operations in the single digitization sample unit.
This information can not be stored in the data flow, since it would imply a doubling of the data size (a further 2~bytes variable would be required to be associated to each data sample). 
Nevertheless, the time lapse can be computed as the difference from the time stamp of the current event and the ``last time tag'' stored in an auxiliary volatile array of data, array which is updated each time a new event is processed. 
Drawback is that this requires to process the events strictly sequentially, and prevent parallel analysis of different events by multiple threads.
To by-pass this problem, the parallelization scheme was moved inside the single event processing, so that multiple threads can process in parallel different chunks of the same event (see figure \ref{simp_fig2}).
Formerly, different events were processed in parallel by different analysis threads. 
Sustainable acquisition rate is 800~Hz (over 1~kHz, disk access limited, without online pre-processing).

\section{The online analysis client} \label{mola}

MOLA (MAGIC Online Analysis) is a multithread C++ program used to obtain on-the-fly estimate of the gamma-ray flux from sources in the FoV of the telescope during MAGIC observations.
The program runs simultaneously with the data acquisition software and acts as receiving client of the event informations computed at the very moment the events are acquired by each telescope.
In fact, as mentioned in the previous section, the DAQ softwares of MAGIC-I and MAGIC-II compute independently the signal and arrival time of each pixel of the telescope cameras.
In this way the calculation of the image parameters and the latest steps of the data analysis are outsourced to an independent program on an independent computer.

The multithread program structure consists of three threads: two \emph{reading} threads and one \emph{analyzing} thread.
The two reading threads are appointed to receive the data stream from the two DAQs asynchronously and perform the non-stereo analysis steps. 
The main analysis thread is instead appointed to match the events from the two streams and perform the stereoscopic reconstruction.

Two independent streams 
are activated once the program start, and each time the observation of a source is finished the current results are stored and the analysis reset.
The event informations are transferred through TCP/IP by means of ASCII \emph{event reports}, which contain all the informations needed for the data analysis. 
The event report format consists of an \emph{header} and a \emph{body}. 
The report header contains, among the others, the following key information:
\begin{itemize}
\item The telescope number and readout state.
\item The source name and sky coordinates.
\item The current data run number.
\item The local DAQ event number.
\item The stereo trigger event number.
\item The event time stamp and trigger pattern.
\end{itemize}

\begin{figure}
\centering
\includegraphics[width=0.19\textwidth]{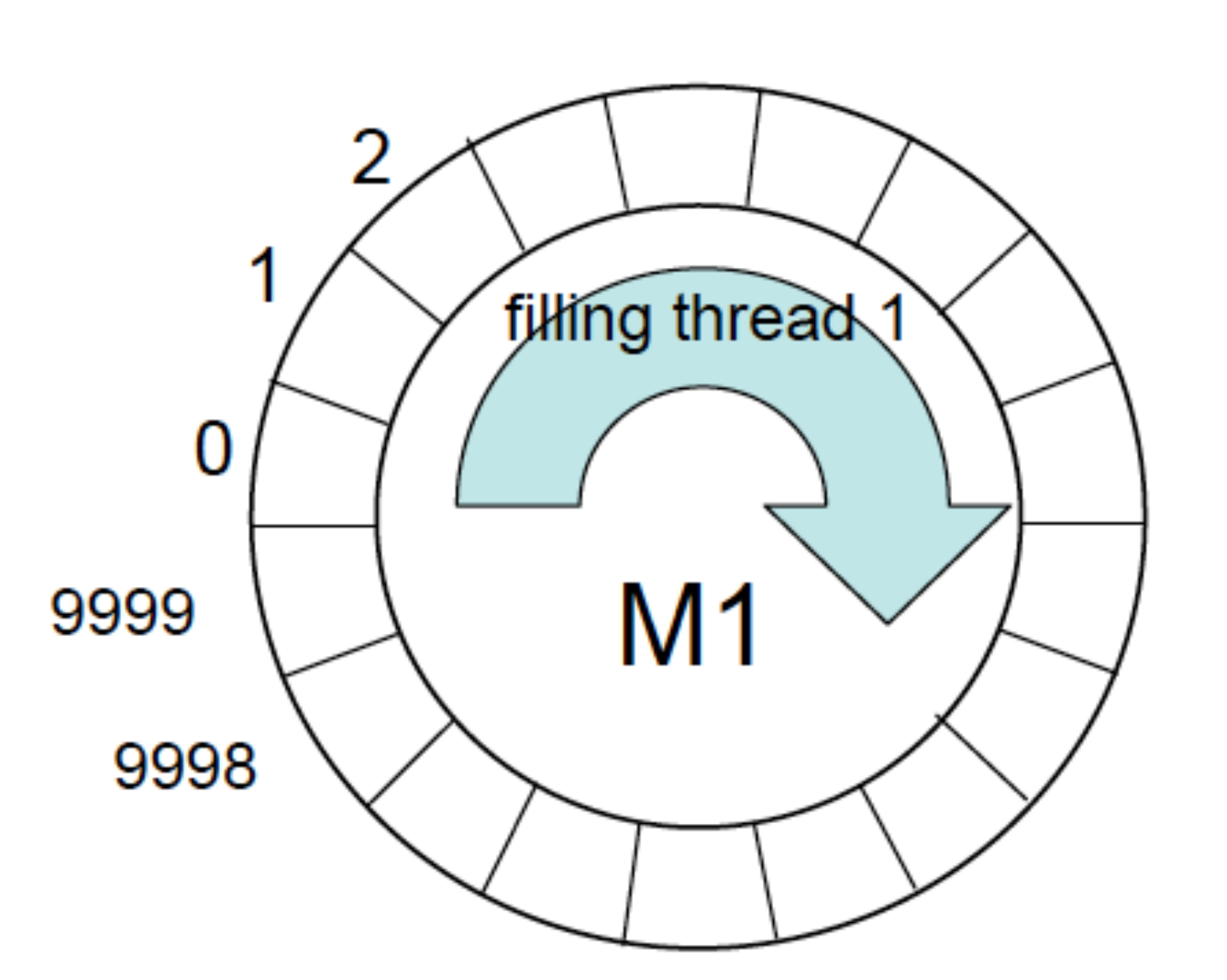}
\includegraphics[width=0.20\textwidth]{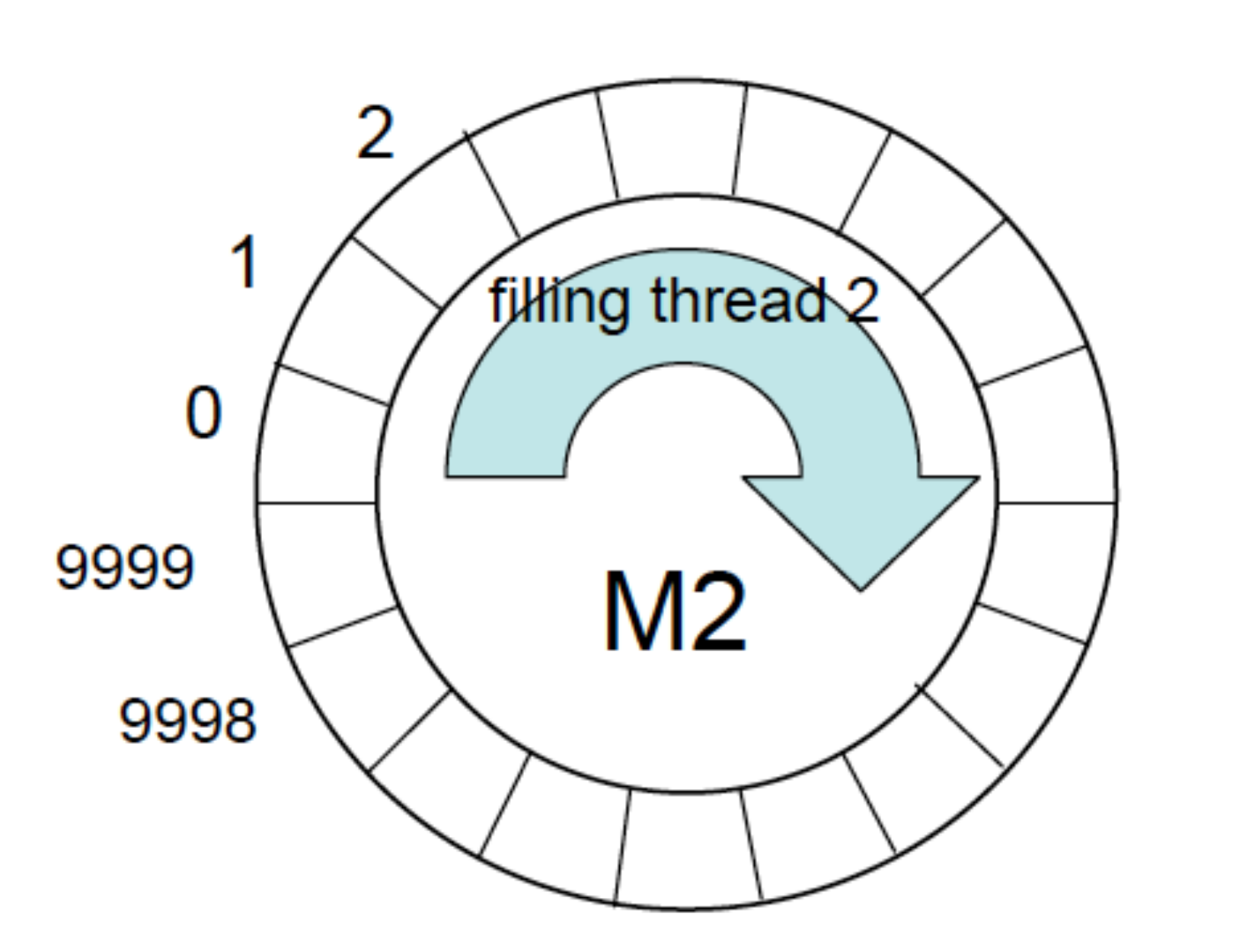}
\caption{The events are received by MOLA via two independent data streams. Events are temporary stored in two ring buffers while a third thread proceeds to search for event matching.}
\label{simp_fig4}
\end{figure}

The report body contains instead the pixel signals (integrated charge), computed with a simple six samples sliding window extractor \cite{bib:perfICRC}, and the signal arrival time with respect to the time window, obtained by the same algorithm. 
The actual data (pixel-wise) are packed in two consecutive lists of number in hexadecimal format to reduce the report size.
The events are first stored in two independent ring buffer, one for each data stream (see figure \ref{simp_fig4}). 
This is done continuously, in order to optimize the data flow and avoid to block the data streams. 
The task of each reading thread are:
\begin{itemize}
\item Receive, interpret and temporarily store the relevant information from the event reports.
\item Relative pixel calibration (flat-fielding) and identification of dead pixels using interleaved calibration events.
\item Check pedestal events to identify and interpolate noisy pixels with low signals (e.g. due to stars).
\item Image cleaning to select pixels with a significant Cherenkov signal.
\item Calculation of image parameters, using standard MAGIC analysis software data structures (see \cite{bib:MARS}).
\end{itemize}

\begin{figure*}[!t]
\centering
\includegraphics[width=.85\textwidth]{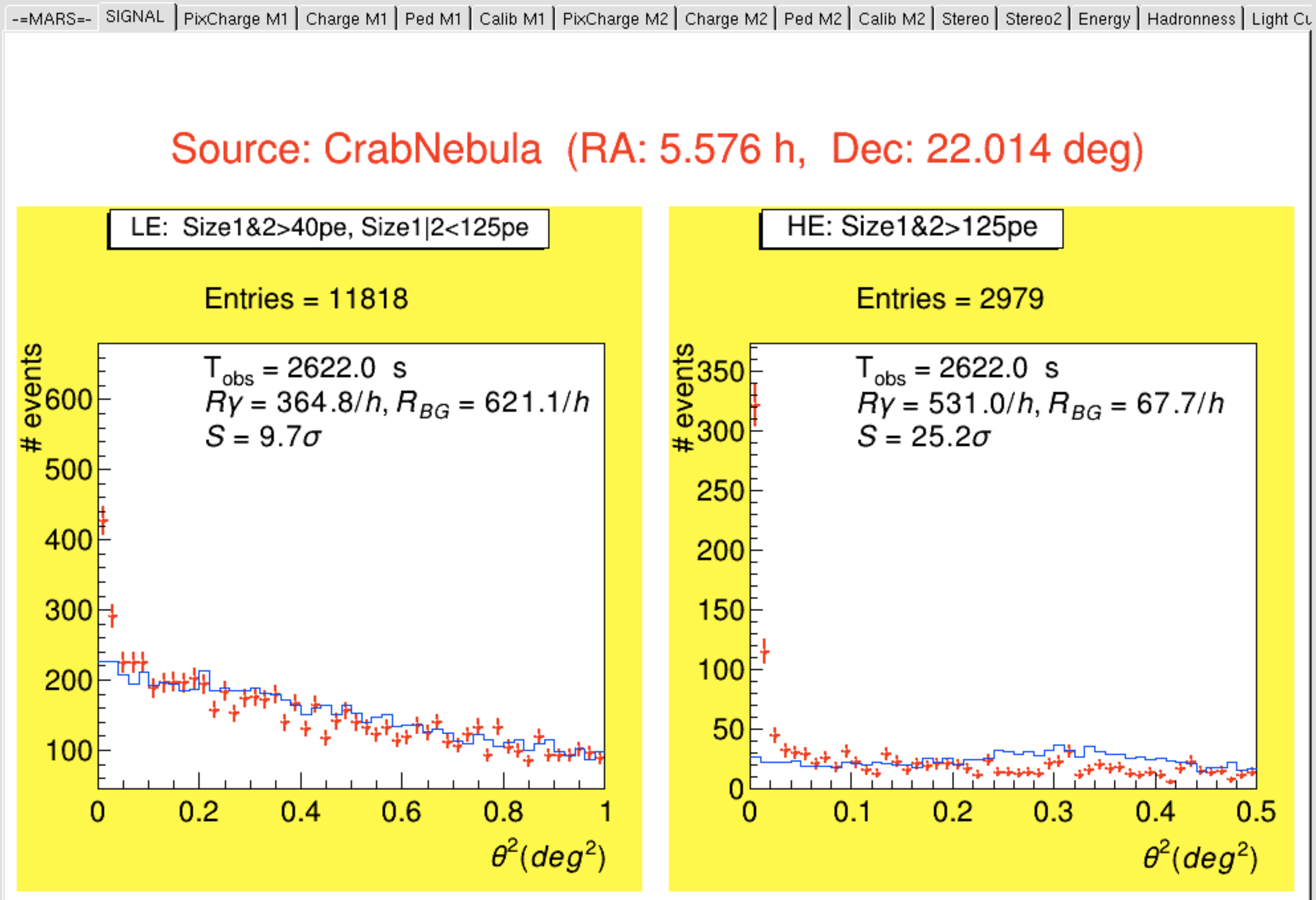}
\caption{Examples of the MOLA program output: $\theta^2$ plots running plots (LE left, HE right) after $\sim$40 minutes of observation of the Crab~Nebula source in December 2012. The panel background becomes yellow if an excess with statistical significance greater than 3~$\sigma$ is detected. Being automatically generated, these results must be considered as \emph{preliminary}.}
\label{wide_fig5}
\end{figure*}

Single telescope events have then to be combined in stereo events in order to exploit the full potential of the stereo imaging technique. 
Stereo events are recognized by means of the \emph{level 3} (L3) trigger number, which is generated by the stereo trigger and is then unique for both telescopes, differently  from the other types of triggers which are acquired\footnote{Interleaved pedestal and calibration triggers are issued, at a rate of 25~Hz respectively, during data data taking.}. 
For each telescope MOLA calculates an estimated shower direction from a set of relevant image parameters (including the time gradient along the major image axis) by using the random forest method \cite{bib:RF}\cite{bib:dispRF}. 
A weighted average of the two estimates provides the final estimate of the event direction.

The task which have to be accomplished in order to obtain high level analysis results are performed by the stereo analysis thread, and can be summarized as follow:
\begin{itemize}
\item Identify matching stereo events by means of the unique L3 trigger number.
\item Calculate the image axesÕ crossing (event direction).
\item Calculate shower core impact point and impact parameters.
\item Apply the background suppression by means of the \emph{hadronness} gamma/hadron likelihood parameter \cite{bib:MARS}.
\item Apply cuts and compute the $\theta^2$ \emph{plot} with respect to the candidate source position.
\item Produce $\gamma$-ray excess sky-maps for LE and HE, where new $\gamma$-ray sources would eventually show up.
\item Produce lightcurves of the measured gamma-ray flux during the current observations. 
\end{itemize}

Results are produced for two energy ranges: Low Energy (LE) and High Energy (HE) depending of the size of the event image in phe.  
In the HE range, the sensitivity of the MOLA analysis have been estimated in 1.4\% of the Crab~Nebula flux in 50~h observation time.

MOLA provides to the telescope operators all the necessary information to judge the behavior of the aimed astrophysical sources like $\theta^2$ plots and excess sky-map, together with diagnostic information related to the signal calibration and the image parameters calculation.
A 600~Hz data transfer and data analysis rate (three times the typical stereo acquisition rate) have been tested on site in La Palma.


\section{Conclusions}

After the 2011-2012 major upgrade, both MAGIC telescopes are equipped with the same upgraded readout system. 
In particular, the data acquisition software part is now identical for both systems, fact which significantly simplifies its maintenance.
It is a quite compact system, of one computer per telescope only, but nevertheless well performing given the sustainable data acquisition rate of ~800~Hz, with full data pre-processing. 

Moreover, the basic DAQ data analysis, formerly used only for data check monitoring, is now successfully exploited jointly with MOLA, the online analysis client for real-time monitoring of $\gamma$-rays emission during data taking.
The interaction between the DAQs and MOLA is designed such that the data acquisition has priority over the online data analysis. 
This means that in case the DAQs are taking data at a rate above the data transfer limit, some events are simply not sent and data acquisition is not slowed down.
Online analysis up to 600~Hz acquisition rate is possible, with no event loss in the data transfer. 

Together, the DAQs and the MOLA online analysis programs allow an immediate feedback in case of a rapid gamma-ray flare of an observed astrophysical source, as already been proved by the IC~310 flare in November 2012 or the huge flare of Mrk~421 in April 2013.

\vspace*{0.5cm}
\footnotesize{{\bf Acknowledgment: }{D. Tescaro acknowledges financial support from the Spanish Ministry of Economy and Competitiveness (MINECO) under the 2011 Severo Ochoa Program MINECO SEV-2011-0187. }}

\end{document}